\documentclass[preprint2]{aastex7}
\usepackage{textcomp}
\usepackage{multirow}
\usepackage{hyphenat}
\graphicspath{{./}{figures/}}

\DeclareTextSymbolDefault{\dh}{T1}
\shortauthors{Sandoval et al.}
\shorttitle{TOI-5349b: A Saturn-like GEM}
\begin{document}

\title{\textit{Searching for GEMS:} TOI-5349b is a Saturn-like planet orbiting a metal-rich early M-dwarf}

\correspondingauthor{Angeli Sandoval}
\email{angelis0821@gmail.com}

\author[0000-0003-1133-1027]{Angeli Sandoval}
\affiliation{The Graduate Center of the City University of New York, 365 Fifth Avenue, New York, NY 10016, USA}
\email{angelis0821@gmail.com}

\author[0000-0003-4835-0619]{Caleb I. Ca\~nas}
\altaffiliation{NASA Postdoctoral Fellow}
\affiliation{NASA Goddard Space Flight Center, 8800 Greenbelt Road, Greenbelt, MD 20771, USA}
\email{c.canas@nasa.gov}

\author[0000-0001-8401-4300]{Shubham Kanodia}
\affil{Earth and Planets Laboratory, Carnegie Institution for Science, 5241 Broad Branch Road, NW, Washington, DC 20015, USA}
\email{skanodia@carnegiescience.edu}  

\author[0000-0001-8020-7121]{Knicole D. Col\'on}
\affiliation{NASA Goddard Space Flight Center, 8800 Greenbelt Road, Greenbelt, MD 20771, USA}
\email{knicole.colon@nasa.gov}

\author[0000-0002-0048-2586]{Andrew Monson}
\affil{Steward Observatory, The University of Arizona, 933 N.\ Cherry Avenue, Tucson, AZ 85721, USA}
\email{andymonson@arizona.edu}

\author[0000-0002-2401-8411]{Alexander Larsen}
\affiliation{Department of Physics \& Astronomy, University of Wyoming, Laramie, WY 82070, USA}
\email{alarse15@uwyo.edu}

\author[0000-0002-5817-202X]{Tera N. Swaby}
\affiliation{Department of Physics \& Astronomy, University of Wyoming, Laramie, WY 82070, USA}
\email{tswaby@uwyo.edu}

\author[0000-0002-4475-4176]{Henry A. Kobulnicky}
\affiliation{Department of Physics \& Astronomy, University of Wyoming, Laramie, WY 82070, USA}
\email{ChipK@uwyo.edu}

\author{Philip I. Choi}
\affiliation{Pomona College, 333 N. College Way Claremont, CA 91711, USA}
\email{Philip.Choi@pomona.edu}

\author[0000-0002-8623-8268]{Sage Santomenna}
\affiliation{Pomona College, 333 N. College Way Claremont, CA 91711, USA}
\email{mstp2022@mymail.pomona.edu}

\author{Pei Qin}
\affiliation{Pomona College, 333 N. College Way Claremont, CA 91711, USA}
\email{pqaa2018@mymail.pomona.edu}

\author[0009-0009-4977-1010]{Michael Rodruck}
\affil{Department of Physics, Engineering, and Astrophysics, Randolph-Macon College, Ashland, VA 23005, USA}
\email{michaelrodruck@rmc.edu}

\author[0000-0001-9662-3496]{William D. Cochran}
\affil{McDonald Observatory and Center for Planetary Systems Habitability, The University of Texas at Austin, Austin, TX 78730, USA}
\email{wdc@astro.as.utexas.edu}

\author[0009-0003-1142-292X]{Nina Brown}
\affil{Department of Astronomy \& Astrophysics, University of Chicago, Chicago, IL 60637, USA}
\email{ninabrown@uchicago.edu}

\author[0000-0003-2404-2427]{Madison Brady}
\affil{Department of Astronomy \& Astrophysics, University of Chicago, Chicago, IL 60637, USA}
\email{mtbrady@uchicago.edu}

\author[0000-0003-4526-3747]{Andreas Seifahrt}
\affil{Department of Astronomy \& Astrophysics, University of Chicago, Chicago, IL 60637, USA}
\email{seifahrt@uchicago.edu}

\author[0000-0002-5463-9980]{Arvind F. Gupta}
\affiliation{U.S. National Science Foundation National Optical-Infrared Astronomy Research Laboratory, 950 N.\ Cherry Ave., Tucson, AZ 85719, USA}
\email{arvind.gupta@noirlab.edu}

\author[0000-0002-3985-8528]{Jesus Higuera}
\affiliation{U.S. National Science Foundation National Optical-Infrared Astronomy Research Laboratory, 950 N.\ Cherry Ave., Tucson, AZ 85719, USA}
\email{jesus.higuera@noirlab.edu}

\author[0000-0002-0885-7215]{Mark E. Everett}
\affiliation{U.S. National Science Foundation National Optical-Infrared Astronomy Research Laboratory, 950 N.\ Cherry Ave., Tucson, AZ 85719, USA}
\email{mark.everett@noirlab.edu}

\author[0009-0000-2465-3119]{Zuri Barksdale}
\affiliation{Department of Physics and Astronomy, Howard University, Washington, DC 20059, USA}
\email{Zuri.Barksdale@bison.howard.edu}

\author[0000-0003-4508-2436]{Ritvik Basant}
\affil{Department of Astronomy \& Astrophysics, University of Chicago, Chicago, IL 60637, USA}
\email{rbasant@uchicago.edu}

\author[0000-0003-4733-6532]{Jacob L.\ Bean}
\affil{Department of Astronomy \& Astrophysics, University of Chicago, Chicago, IL 60637, USA}
\email{jacobbean@uchicago.edu}

\author[0000-0002-2144-0764]{Scott A. Diddams}
\affiliation{Electrical, Computer \& Energy Engineering, University of Colorado, 425 UCB, Boulder, CO 80309, USA}
\affiliation{Department of Physics, University of Colorado, 2000 Colorado Avenue, Boulder, CO 80309, USA}
\email{scott.diddams@colorado.edu}

\author[0000-0001-6340-8220]{Giannina Guzm\'an Caloca}
\affiliation{Department of Astronomy, University of Maryland, College Park, MD 20742, USA}
\affiliation{NASA Goddard Space Flight Center, 8800 Greenbelt Road, Greenbelt, MD 20771, USA}
\email{gguzmanc@umd.edu}

\author[0000-0003-1312-9391]{Samuel Halverson}
\affiliation{Jet Propulsion Laboratory, California Institute of Technology, 4800 Oak Grove Drive, Pasadena, California 91109}
\email{samuel.halverson@jpl.nasa.gov}

\author[0000-0002-2990-7613]{Jessica Libby-Roberts}
\affil{Department of Astronomy \& Astrophysics, 525 Davey Laboratory, The Pennsylvania State University, University Park, PA 16802, USA}
\affil{Center for Exoplanets and Habitable Worlds, 525 Davey Laboratory, The Pennsylvania State University, University Park, PA 16802, USA}
\email{jer5346@psu.edu}

\author[0000-0002-9082-6337]{Andrea S.J. Lin}
\affiliation{Department of Astronomy, California Institute of Technology, 1200 E California Blvd, Pasadena, CA 91125, USA}
\email{asjlin@caltech.edu}

\author[0000-0002-4671-2957]{Rafael Luque}
\affil{Department of Astronomy \& Astrophysics, University of Chicago, Chicago, IL 60637, USA}
\email{rluque@uchicago.edu}
\affil{NHFP Sagan Fellow}

\author[0000-0001-8127-5775]{Arpita Roy}
\affiliation{Astrophysics \& Space Institute, Schmidt Sciences, New York, NY 10011, USA}
\email{arpita308@gmail.com}


\author[0000-0001-7409-5688]{Gu\dh mundur Stef\'ansson}
\affil{Anton Pannekoek Institute for Astronomy, University of Amsterdam, Science Park 904, 1098 XH Amsterdam, The Netherlands} 
\email{g.k.stefansson@uva.nl}



\begin{abstract}
We report the confirmation and analysis of TOI-5349b, a transiting, warm, Saturn-like planet orbiting an early M-dwarf with a period of $\sim$3.3~days, which we confirmed as part of the \textit{Searching for GEMS} (Giant Exoplanets around M-dwarf Stars) survey. TOI-5349b was initially identified in photometry from NASA's \textit{Transiting Exoplanet Survey Satellite} (TESS) mission and subsequently confirmed using high-precision radial velocity (RV) measurements from the Habitable-zone Planet Finder (HPF) and MAROON-X spectrographs, and from ground-based transit observations obtained using the 0.6-m telescope at Red Buttes Observatory (RBO) and the 1.0-m telescope at the Table Mountain Facility of Pomona College. From a joint fit of the RV and photometric data, we determine the planet's mass and radius to be $0.40\pm 0.02~\mathrm{M_J}$ ($127.4_{-5.7}^{+5.9}~M_\oplus$) and $0.91\pm 0.02~\mathrm{R_J}$ ($10.2\pm 0.3~R_\oplus$), respectively, resulting in a bulk density of $\rho_p=0.66 \pm0.06~\mathrm{g~cm^{-3}}$ ($\sim0.96$ the density of Saturn). We determine that the host star is a metal-rich M1-type dwarf with a mass and radius of $0.61 \pm 0.02~M_\odot$ and $0.58\pm 0.01~R_\odot$, and an effective temperature of $T_\mathrm{eff} = 3751 \pm 59$~K. Our analysis highlights an emerging pattern, exemplified by TOI-5349, in which transiting GEMS often have Saturn-like masses and densities and orbit metal-rich stars. With the growing sample of GEMS planets, comparative studies of short-period gas giants orbiting M-dwarfs and Sun-like stars are needed to investigate how metallicity and disk conditions shape the formation and properties of these planets.
\end{abstract}



\section{Introduction} \label{sec:intro}
Since its launch in 2018, NASA's \textit{Transiting Exoplanet Survey Satellite} \citep[TESS;][]{Ricker2015} has enabled the discovery of a wide range of exoplanets, including tens of giant planets around M-dwarfs so far, greatly expanding the previous population of these systems. These Giant Exoplanets around M-dwarf Stars (GEMS), which are typically defined by host stars with effective temperatures $T_\mathrm{eff} \lesssim 4000$~K, and radii $8~R_\oplus \leq R_p \lesssim 15~R_\oplus$ or for the non-transiting ones, minimum masses $M_p \sin i \gtrsim 80~M_\oplus$ \citep{Kanodia2024a}. \textit{The Searching for GEMS} survey \citep{Kanodia2024} is designed to explore the formation pathways of these gas giants around M-dwarfs by constraining their system properties, orbital architectures, and comparing them with similar systems.

GEMS are of particular interest because the presence of massive gas giants in low-mass stellar environments challenges traditional planet formation models for core accretion. The prediction from theory is that gas giants are less likely to form \textit{in-situ} (at their observed location) around low-mass stars due to reduced disk masses and longer core growth timescales \citep{Laughlin2004}. Observations of planets orbiting FGK dwarfs \citep{Lin1996} and M-dwarfs \citep[e.g.,][]{Kanodia2023,Hobson2023,BossKanodia2023} propose that these planets formed at larger orbital separations through core accretion or gravitational instability and later migrated inward to their current orbits. 

As part of the GEMS survey, we present the confirmation and characterization of TOI-5349b, a Saturn-like planet ($M_p = 0.40~M_\mathrm{J}$, $R_p=~0.91~R_\mathrm{J}$, $\rho_p = 0.66$~g~cm$^{-3}$) orbiting an early M-dwarf. We confirm the planetary nature of TOI-5349b using TESS photometry, ground-based photometry from the Red Buttes Observatory and the Table Mountain Facility of Pomona College, as well as high-precision radial velocities (RVs) from the Habitable-zone Planet Finder (HPF) and MAROON-X spectrographs. In Section~\ref{sec:obs}, we describe in more detail the photometric and radial velocity observations. In Section~\ref{sec:stellar}, we detail the characterization of the host star. Section~\ref{sec:datafit} presents the joint modeling of the RV and photometric data. We discuss the implications of TOI-5349b in the context of the GEMS sample in Section~\ref{sec:discussion}, and summarize our conclusions in Section~\ref{sec:conclusion}.

\section{Observations}
\label{sec:obs}
\subsection{Photometry}
\subsubsection{TESS}
\begin{figure*}[!ht]
\epsscale{1.15}
\plotone{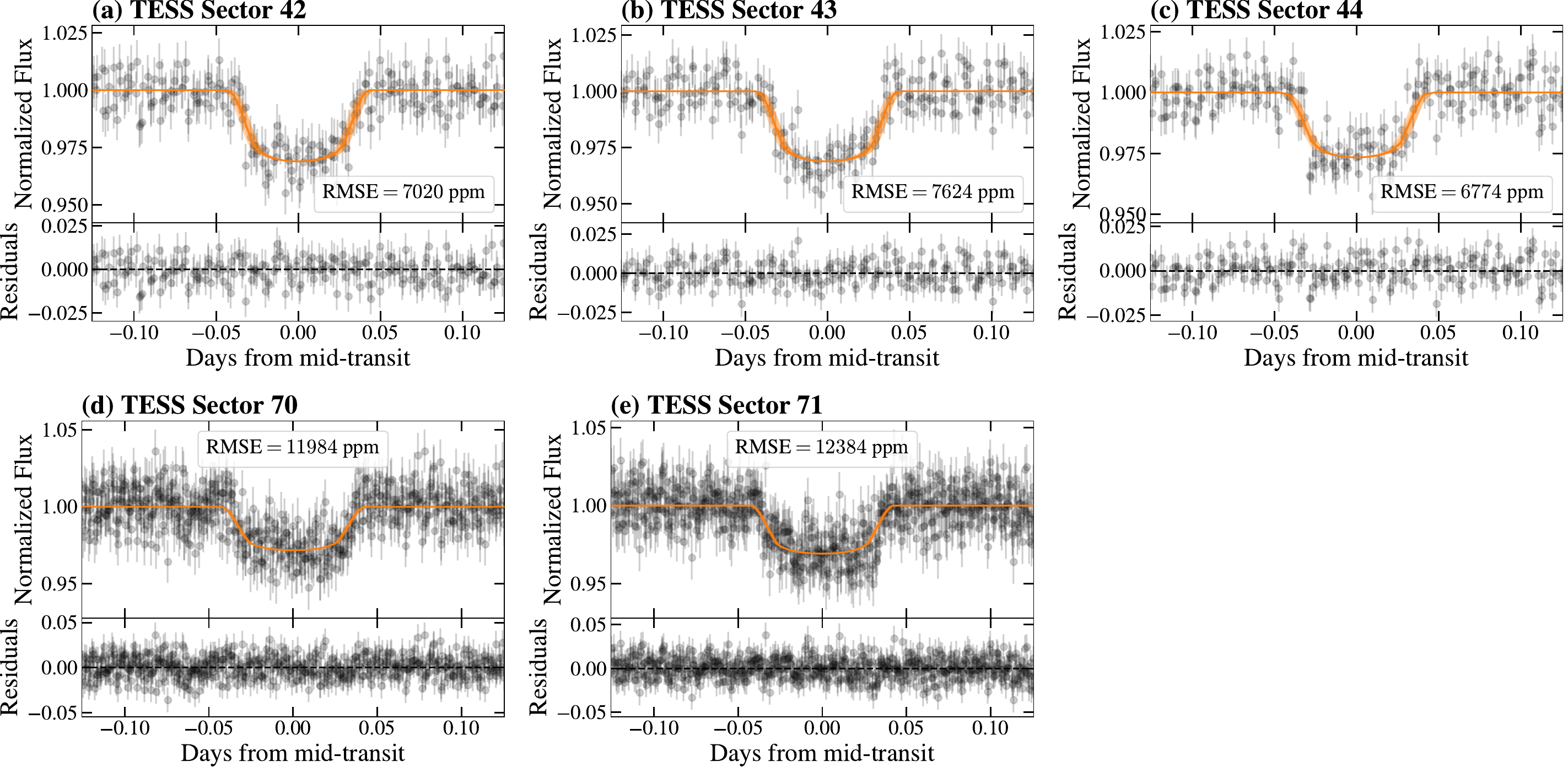}
\caption{\textbf{(a)-(e)} Median normalized, phase-folded TESS light curves for various sectors of TOI-5349 derived with \texttt{tglc}. The solid line is the best-fitting transit model. In each panel, the shaded regions are the \(1\sigma\) (darkest), \(2\sigma\), and \(3\sigma\) (lightest) extent of the model posteriors. The root mean square error (RMSE) is indicated for reference. See Section \ref{sec:datafit} for a detailed description of the modeling.}
\label{fig:tesslc}
\end{figure*}

TOI-5349.01 was first identified as a candidate using photometric observations from TESS as part of the faint star search for planet candidates around stars with TESS magnitude $T>12$ \citep{Kunimoto2022}. The NASA TESS mission \citep{Ricker2015} first observed TOI-5349 (TIC 26054627; Gaia DR3 58372904816938240) in Sector 42 (2021 August 20 -- 2021 September 16), 43 (2021 September 16-- 2021 October 12) and 44 (2021 October 12 -- 2021 November 6) at a 600 s cadence. The target was re-observed in Sectors 70 (2023 September 20 - 2023 October 16) and 71 (2023 October 16 -- 2023 November 11) at a 200 s cadence. Similar to \cite{Kanodia2024a}, we extracted the light curves from the TESS full-frame images (FFIs) for each sector using the TESS-Gaia Light Curve (\texttt{tglc}\footnote{\url{https://github.com/TeHanHunter/TESS_Gaia_Light_Curve}}) pipeline \citep{Han2023}, which uses photometry from Gaia DR3 as priors to remove contamination from background stars. We used a $90\times90$ pixel cutout centered on TOI-5349 to sample the point-spread function and derive the contamination-corrected light curves. In this work, we analyzed the calibrated aperture light curves (\texttt{cal\_aper\_flux}) that are derived using a $3\times3$ aperture centered on TOI-5349. The phase-folded photometry is shown in \autoref{fig:tesslc}.

\subsubsection{Red Buttes Observatory 0.6-m telescope}
\begin{figure*}[!ht]
\epsscale{1.15}
\plotone{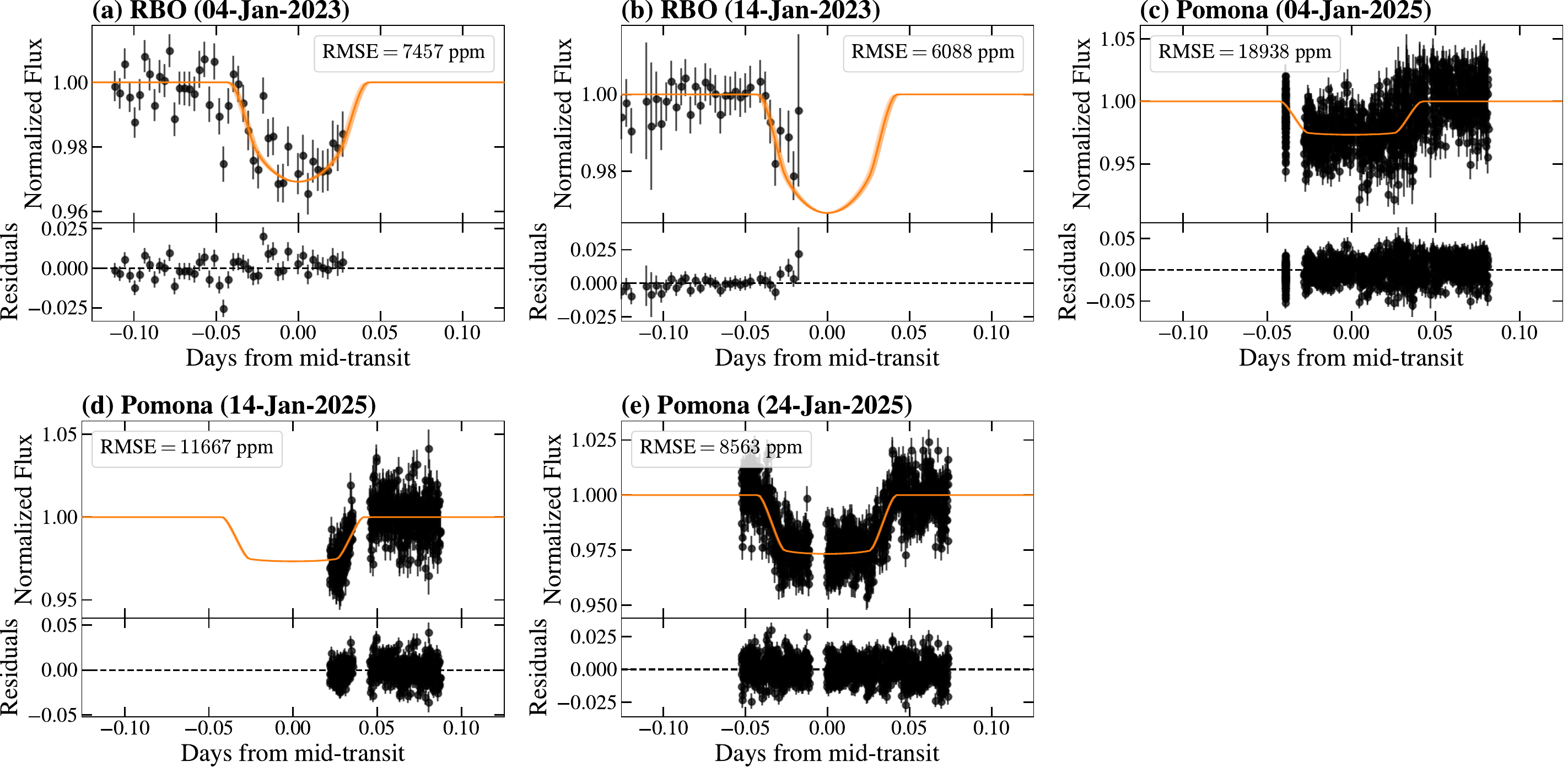}
\caption{\textbf{(a)-(e)} Identical to \autoref{fig:tesslc} but for ground-based RBO (Bessel I) and Pomona (Sloan $i^\prime$) data.}
\label{fig:lc}
\end{figure*}

We obtained two ground-based transit observations of TOI-5349 on 2023 January 4, and 2023 January 14, using the 0.6-m telescope at Red Buttes Observatory \citep{Kasper2016} in Wyoming, USA. The transits were observed with the AltaF16 camera, employing a Bessel I filter and a plate scale of 0.731\arcsec/pixel. Observations were performed with 2$\times$2 on-chip pixel binning at a 240 s cadence. Weather and seeing conditions resulted in partial transits on both nights with RBO. The light curves of both visits were derived using differential photometry derived with \texttt{AstroImageJ} \citep{Collins2017} in which we adopted an aperture radius of 7 pixels (5.1\arcsec{}), inner sky radius of 12 pixels (8.8\arcsec{}) and an outer sky radius of 17 pixels (12.4\arcsec{}). The RBO photometry is shown in \autoref{fig:lc}.

\subsection{1.0-m Table Mountain Facility of Pomona College}
We also observed three transits (2025 January 3, 2025 January 14 and 2025 January 23) of TOI-5349 with Pomona College's 1.0-m telescope located at NASA Jet Propulsion Laboratory's Table Mountain Facility in California, USA. The visibility from Table Mountain only allowed for an observation of partial transits (mid-transit through egress) on 2025 3 January and 14 January. The telescope was operated with 1$\times$ 1 binning, a gain of 0.8 $e^-$/ADU, and a plate scale of 0.232\arcsec/pixel. All observations used a 10-second exposure time and were obtained in the Sloan $i^\prime$ filter. Light curves for all visits were extracted using \texttt{AstroImageJ}, using an aperture radius of 17 pixels (3.9\arcsec), an inner sky radius of 29 pixels (6.7\arcsec), and an outer sky radius of 44 pixels (10.2\arcsec). The Pomona photometry is shown in \autoref{fig:lc}.

\subsection{High-resolution spectroscopy}
\subsubsection{HPF}
\begin{figure*}[!ht]
\epsscale{1.15}
\plotone{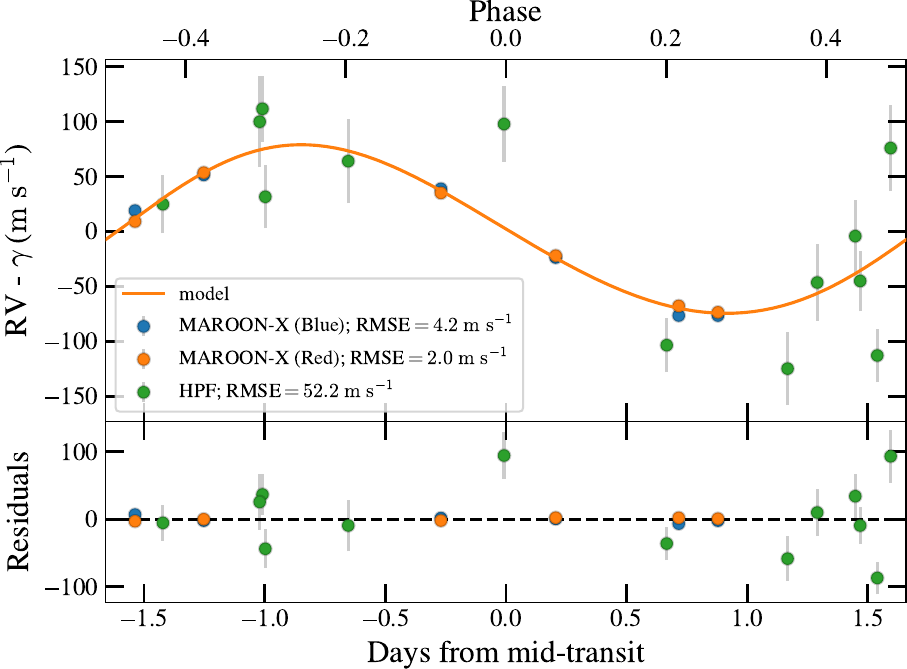}
\caption{\textbf{Top:} Phase-folded RVs for TOI-5349 derived with \texttt{SERVAL}. The best-fitting Keplerian model is denoted with a solid line. The shaded regions denote the \(1\sigma\) (darkest), \(2\sigma\), and \(3\sigma\) (lightest) extent of the model posteriors. The uncertainties for MAROON-X are too small to visualize at the scale shown. \textbf{Bottom:} Phase-folded residuals to the fit. The modeling is described in Section \ref{sec:datafit}.}
\label{fig:rv}
\end{figure*}

Between 2022 October 22 -- 2024 January 29, we obtained 13 visits of TOI-5349 with the Habitable-zone Planet Finder (HPF), a high-resolution, near-infrared ($8080-12780$ \AA) spectrograph ($R \sim 55,000$) that is installed on the 10-m Hobby-Eberly Telescope located at the McDonald Observatory in Texas, USA \citep{Ramsey1998, Hill2021}. HPF is a fiber-fed \citep{Kanodia2018}, thermally stabilized \citep{Stefansson2016}, spectrometer that was designed to obtain high-precision RVs in the near-infrared \citep{Suvrath2012, Suvrath2014}. Each visit consisted of two exposures of 945 seconds each and were executed in a queue operated by HET resident astronomers \citep{Shetrone2007}. The raw spectral data are processed using the \texttt{HXRGproc}\footnote{\url{https://github.com/indiajoe/HxRGproc}} package following the methods described in \cite{Ninan2018}. We performed barycentric correction utilizing the \texttt{barycorrpy}\footnote{\url{https://github.com/shbhuk/barycorrpy}} package \citep{Kanodia2018bc} developed from the algorithms in \citet{WrightEastman2014}. The RV measurements for each 945 s exposure are then derived from the spectra using \texttt{HPF-SERVAL}\footnote{\url{https://github.com/gummiks/hpfserval_lhs3154.git}} \citep{Stefansson2023}, a modified version of the template matching \texttt{SpEctrum Radial Velocity AnaLyser} (\texttt{SERVAL}) framework \citep{Zechmeister2018}. The HPF data have a median per-pixel signal-to-noise (S/N) per unbinned exposure (of 945 s) at 1070 nm of 32. The binned RVs (weighted average of the two exposures) and the respective uncertainties are listed in \autoref{tab:rvs} and shown in \autoref{fig:rv}.

\subsubsection{MAROON-X}
We obtained six visits of TOI-5349 between 2023 October 14 and October 28 (Program GN-2023B-Q-104, PI Kanodia) using MAROON-X, a high-resolution ($R\sim 85,000$), red-optical spectrograph on the 8.1-m Gemini North telescope \citep{Seifahrt2016,Seifahrt2018,Seifahrt2020,Seifahrt2022}. MAROON-X has two channels with different wavelength coverage consisting of a ``blue'' channel that spans $5000-6700$ \AA{} and a ``red'' channel that covers $6500-9000$ \AA{}. All visits used a fixed exposure time of 1800 s, which provided a median peak S/N of 28.5 and 51.5 in the blue and red channels, respectively. Both channels were exposed simultaneously during each visit. We reduced the data using a custom pipeline originally developed for the CRIRES instrument as outlined in \citep{Bean2010} and measured the RVs for each channel using a modified version of the \texttt{SERVAL} algorithm \citep{Zechmeister2018}. The derived RVs and the respective uncertainties for each channel are listed in \autoref{tab:rvs} and shown in \autoref{fig:rv}. Each channel is considered a separate instrument when fitting the data (see Section \ref{sec:datafit} for details).  
\begin{figure*}
    \centering
    \includegraphics[width=\linewidth]{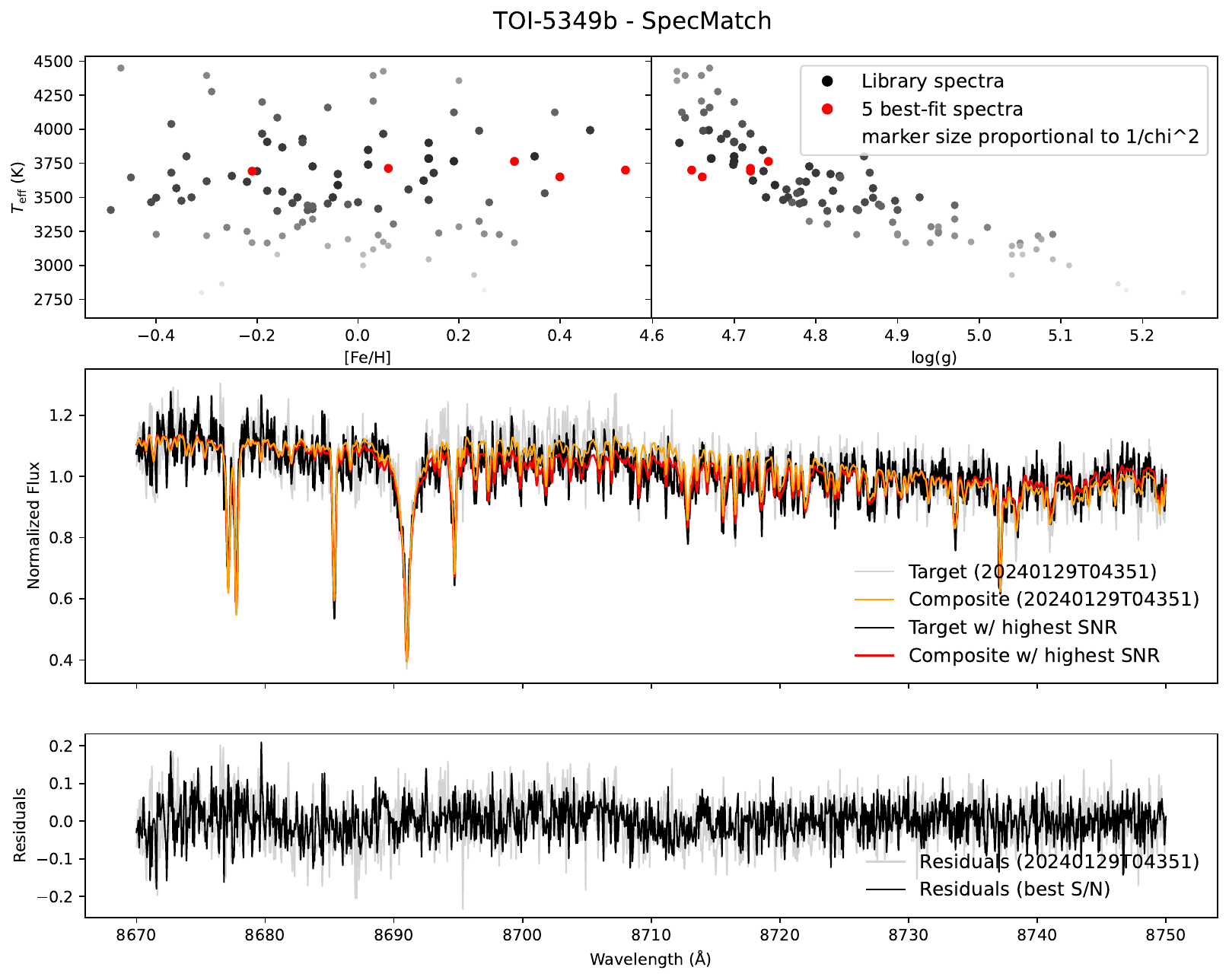}
    \caption{TOI-5349's HPF--SpecMatch spectral fit results for order index 5. Top: two panels show the five best-fit library stars in red that were selected to construct the composite spectrum of TOI-5349b. The size and shading of both the best-fit stars (red) and the remaining library stars (black) scale inversely with the initial $\chi^{2}$ value, where larger and darker points represent better matches to the target spectrum. Middle: the HPF order 5 spectrum compared to its best-fit composite (orange). Bottom: residuals between the spectrum and best-fit composite (black).
}
\label{fig:specmatch-plot}
\end{figure*}

\begin{deluxetable}{cccc}
\tablecaption{RVs for TOI-5349. \label{tab:rvs}}
\tablehead{
\colhead{Time}&
\colhead{RV} &
\colhead{$\sigma$}&
\colhead{Instrument} \\
\colhead{$\mathrm{BJD_{TDB}}$}&
\colhead{$\mathrm{m~s^{-1}}$} &
\colhead{$\mathrm{m~s^{-1}}$} &
\colhead{}
}
\startdata
2459874.965870 &   64.83 &  32.89 &             HPF \\
2459884.939955 &   23.96 &  27.70 &             HPF \\
2459897.910783 &  $-55.94$ &  33.24 &             HPF \\
2459915.642095 &  180.81 &  29.95 &             HPF \\
2459916.643227 &  166.92 &  34.24 &             HPF \\
2459922.633343 &  133.12 &  38.43 &             HPF \\
2459924.825907 &  $-43.96$ &  24.41 &             HPF \\
2459954.742606 &  145.06 &  39.35 &             HPF \\
2459967.709932 &   22.55 &  35.04 &             HPF \\
2459978.670997 &  169.07 &  41.28 &             HPF \\
2459988.648056 &  100.66 &  28.67 &             HPF \\
2460223.796071 &   93.89 &  26.66 &             HPF \\
2460338.693217 &  $-34.54$ &  24.50 &             HPF \\
\hline
2460232.059785 &   $-5.09$ &   3.87 & MAROON-X (Blue) \\
2460234.901990 &   57.58 &   4.72 & MAROON-X (Blue) \\
2460235.887519 &  $-57.93$ &   4.78 & MAROON-X (Blue) \\
2460236.952051 &   37.69 &   3.67 & MAROON-X (Blue) \\
2460243.872681 &   70.20 &   4.93 & MAROON-X (Blue) \\
2460246.003767 &  $-57.95$ &   3.36 & MAROON-X (Blue) \\
\hline
2460232.059785 &   $-3.33$ &   3.22 &  MAROON-X (Red) \\
2460234.901990 &   53.81 &   3.84 &  MAROON-X (Red) \\
2460235.887519 &  $-49.14$ &   3.91 &  MAROON-X (Red) \\
2460236.952051 &   27.83 &   3.03 &  MAROON-X (Red) \\
2460243.872681 &   72.33 &   3.96 &  MAROON-X (Red) \\
2460246.003767 &  $-54.91$ &   2.72 &  MAROON-X (Red) \\
\enddata
\tablecomments{The reported HPF RVs are the weighted average of the RVs for measured from the individual 945s exposure.}
\end{deluxetable}

\subsection{Excluding an eclipsing binary scenario with spectroscopy}
We note that since our photometry does not include multiple band passes to vet our object for eclipsing binaries, we relied on high-resolution spectroscopy to exclude this possibility. Our RV time series with HPF began prior to any ground-based transit observations, allowing us to test for stellar companions at the transit period. The HPF RV data did not show evidence of large-amplitude variations or secondary star light that would signify a binary. We further examined the differential line width and the chromatic RV index calculated by \texttt{HPF-SERVAL} which are proxies for line deformations that may be due to binarity or stellar activity \citep[see \S4 in][]{Zechmeister2018}.  These \texttt{HPF-SERVAL} diagnostics displayed no significant variations or correlations with our observed RVs, suggesting that there were no significant line profile distortions in our HPF spectra.

Subsequent MAROON-X observations, obtained at a higher signal-to-noise ratio, confirmed these findings, where the measured RV signal was consistent with a planetary companion and not a binary signal and the MAROON-X differential line width values likewise showed no correlation with the RV data. We therefore conclude that these spectroscopic diagnostics exclude the possibility of an eclipsing binary in the confirmation of TOI-5349b. Furthermore, after we modeled the photometry and RVs (Section \autoref{sec:datafit}), we confirmed that our measured stellar density was identical to the density derived from the transit \citep[within $1\sigma$; see][]{Seager2003,Winn2010, Kipping2014} which suggested TOI-5349b was indeed a planet orbiting this M-dwarf.



\subsection{Speckle imaging with NESSI}
\begin{figure}
    \centering
    \includegraphics[width=\linewidth]{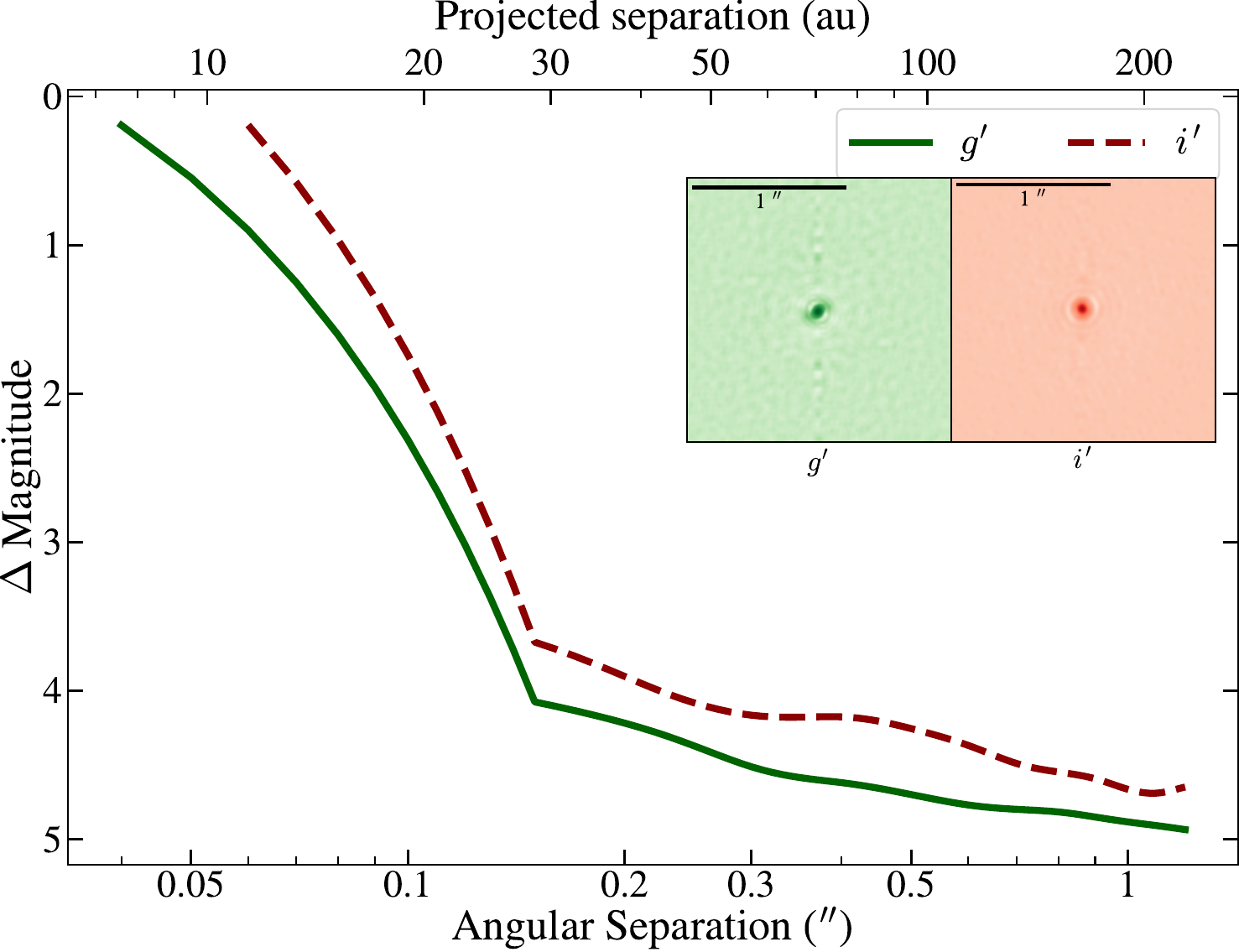}
    \caption{The \(5\sigma\) contrast curves for TOI-5349 obtained with NESSI. The data reveal no bright companions at separations of $0.15\arcsec{}-1.2\arcsec{}$. The insets are the reconstructed images centered on TOI-5349 in the Sloan $g^\prime$ and Sloan $i^\prime$ filters.}
    \label{fig:speckle}
 \end{figure}

We observed TOI-5349 on the night of 2022 September 18 using the NN-Explore Exoplanet Stellar Speckle Imager \citep[NESSI;][]{Scott2018}, a dual-channel speckle imager on the WIYN 3.5-m Telescope at Kitt Peak National Observatory (PI Gupta, 2022B-936991). TOI-5349 was observed in both the Sloan $g^\prime$ and $i^\prime$ filters and the images were reconstructed following the procedures described in \cite{Howell2011}. \autoref{fig:speckle} presents the $5\sigma$ contrast curves, which reveal that there are no bright ($\Delta i^\prime < 3.5$ and $\Delta g^\prime < 4.0$) companions at separations of $0.15-1.2\arcsec$ from TOI-5349. To assess whether TOI-5349 is part of a wide-separation binary system, we cross-matched it with the \cite{El-Badry2021, ElBadryData} catalog\footnote{\url{https://zenodo.org/records/4435257}} and found that it is not listed as a resolved wide binary in Gaia DR3, consistent with the Speckle Imager results.

\section{Stellar Characterization}
\label{sec:stellar}
\subsection{Spectroscopic parameters with HPF}\label{sec:specmatch}
We derived the stellar effective temperature ($T_e$), surface gravity ($\log g_\star$), metallicity ([Fe/H]), and rotational broadening ($v\sin i_\star$) from HPF spectra of TOI-5349 using the \texttt{HPF-SpecMatch}\footnote{\url{https://gummiks.github.io/hpfspecmatch/}} package \citep[][]{Stefansson2020}. \texttt{HPF-SpecMatch} follows a similar methodology presented by \cite{Yee2017} to derive spectroscopic parameters using a weighted linear-combination of the five best-matching empirical spectra from a library of well-characterized stars. In this work, the HPF spectral library consisted of 100 stars that span $2700\mathrm{K} \le T_{e} \le 4500~\mathrm{K}$, $4.63<\log g_\star < 5.26$, and $-0.49 < \mathrm{[Fe/H]} < 0.53$. Spectral matching was performed using the region covering $8534-8645$ \AA{} (HPF order index 5) due to minimal telluric contamination and the parameter uncertainties were the standard deviation of the residuals from a leave-one-out cross-validation procedure applied to the library.  

\begin{deluxetable*}{lccc}
\tabletypesize{\scriptsize}
\tablecaption{Summary of stellar parameters for TOI-5349. \label{tab:stellarparam}}
\tablehead{
\colhead{~~~Parameter}& 
\colhead{Description} &
\colhead{Value} &
\colhead{Reference}}
\startdata
\multicolumn{4}{l}{\hspace{-0.2cm} Main identifiers:}  \\
~~~TIC & TESS Input Catalog ID  & 26054627 & TIC \\
~~~TOI & TESS Object of Interest ID  & 5349 &  TSO \\
~~~Gaia DR3 & Gaia Source ID & 58372904816938240 & Gaia DR3 \\
\hline
\multicolumn{4}{l}{\hspace{-0.2cm} Coordinates, proper motion, distance, maximum extinction, and spectral type:} \\
~~~$\alpha_{\mathrm{J2016}}$ &  Right Ascension (RA) & 03:30:51.01 & Gaia DR3 \\
~~~$\delta_{\mathrm{J2016}}$ &  Declination (Deg) & 20:52:46.65 & Gaia DR3 \\
~~~$\mu_{\alpha}$ &  Proper motion (RA, mas yr$^{-1}$) & $37.20 \pm 0.03$ & Gaia DR3 \\
~~~$\mu_{\delta}$ &  Proper motion (Dec, mas yr$^{-1}$) & $-19.56 \pm 0.02$ & Gaia DR3  \\
~~~$l$ &  Galactic longitude & 165.83445 & Gaia DR3 \\
~~~$b$ &  Galactic latitude & -28.41487 & Gaia DR3  \\
~~~$\varpi$ &  Parallax (mas)  & $5.26 \pm 0.03$ & Gaia DR3\\
~~~\(A_{V,max}\) & Maximum visual extinction & 0.04 & Green\\
~~~Spectral type & \nodata & M$1\pm0.5$ & LAMOST \\
\hline
\multicolumn{4}{l}{\hspace{-0.2cm} Broadband photometric magnitudes:}  \\
~~~$g'$ & PS1 $g'$ mag & $16.38\pm0.01$ & PanSTARRS\\
~~~$r'$ & PS1 $r'$ mag & $15.155 \pm 0.004$ & PanSTARRS\\
~~~$i'$ & PS1 $i'$ mag & $14.242 \pm 0.001$ & PanSTARRS\\
~~~$z'$ & PS1 $z'$ mag & $13.84 \pm 0.01$ & PanSTARRS\\
~~~$y'$ & PS1 $y'$ mag & $13.622 \pm 0.008$ & PanSTARRS\\
~~~$T^a$ & TESS $T$ mag & $13.8117 $ & TIC\\
~~~$J$ & 2MASS $J$ mag & $12.43 \pm 0.02$ & 2MASS\\
~~~$H$ & 2MASS $H$ mag & $11.73 \pm 0.02$ & 2MASS\\
~~~$K_s$ & 2MASS $K_s$ mag & $11.51 \pm 0.02$ & 2MASS\\
~~~W1 & WISE1 mag & $11.4 \pm 0.2$ & WISE\\
~~~W2 & WISE2 mag & $11.45 \pm 0.02$ & WISE\\
~~~W3 & WISE3 mag & $11.3 \pm 0.2$ & WISE\\
\hline
\multicolumn{4}{l}{\hspace{-0.2cm} Spectroscopic parameters from \texttt{HPF-SpecMatch}:}\\
~~~$T_{e}$ &  Effective temperature (K) & $3751\pm59$ & This work\\
~~~$\log g_\star$ & Surface gravity (cgs) & $4.72\pm0.04$ & This work\\
~~~$\mathrm{[Fe/H]}$ &  Metallicity (dex) & $0.50\pm0.16$ & This work\\
~~~$v\sin i_{\star}$ & Rotational broadening (km s$^{-1}$) & $<2$ & This work\\
\hline
\multicolumn{4}{l}{\hspace{-0.2cm} Model-dependent parameters from a stellar SED and isochrone fit$^b$:}\\
~~~$M_\star$ &  Mass ($M_{\odot}$) & $0.61 \pm 0.02$ & This work\\
~~~$R_\star$ &  Radius ($R_{\odot}$) & $0.58 \pm 0.01$ & This work\\
~~~$L_\star$ &  Luminosity ($L_{\odot}$) & $0.063\pm0.002$ & This work\\
~~~$\rho_\star$ &  Density ($\mathrm{g~cm^{-3}}$) & $4.4\pm0.3$ & This work\\
~~~$d$ &  Distance (pc) & $190\pm1$ & This work\\
~~~$A_v$ & Visual extinction (mag) & $0.02 \pm 0.01$ & This work\\
~~~Age & Age (Gyrs) & $9_{-4}^{+3}$ & This work\\
\hline
\hline
\multicolumn{4}{l}{\hspace{-0.2cm} Other stellar parameters:}\\
~~~$RV$ & Systemic RV (km s$^{-1}$) & $40.42 \pm 0.05$ & This work\\
~~~$(U, V, W)_{\mathrm{LSR}}$ &  Galactic velocities w.r.t. LSR$^c$ (km s$^{-1}$) & $-36.2 \pm 0.8,~-13.9 \pm 0.5,~-4.7 \pm 0.4$ & This work\\
\enddata
\tablerefs{TIC \citep{Stassun2019}, TSO \citep{Guerrero2021}, Gaia DR3 \citep{GaiaCollaboration2022}, Green \citep{Green2019}, LAMOST \citep{Zhong2019}, PanSTARRS \citep{Chambers2016}, 2MASS \citep{Cutri2003}, WISE \citep{Wright2010}}
\tablenotetext{a}{TESS is not used in the SED fit.}
\tablenotetext{b}{Derived with the \texttt{EXOFASTv2} package using MIST isochrones.}
\tablenotetext{c}{Calculated using the solar velocities from \cite{Schoenrich2010}.}
\end{deluxetable*}

\autoref{tab:stellarparam} presents the derived spectroscopic parameters with their uncertainties.  TOI-5349 is determined to have $T_{e}=3751\pm59$ K, $\log g_\star=4.72\pm0.05$, and $\mathrm{[Fe/H]=0.50\pm0.16}$. The resolution limit of HPF ($R\sim55,000$) can only place an upper limit of $v \sin i < 2 \mathrm{~km~s^{-1}}$. Similar to the previous results from the \textit{Searching for GEMS} survey \citep{Bernabo2024,Kanodia2024a,Reji2025}, we note that the complexities of M-dwarf spectra and limitations in the \texttt{HPF-SpecMatch} algorithm limit the accuracy of the derived metallicity. We advise caution in interpreting the [Fe/H] beyond a qualitative indicator that TOI-5349 is super-solar in metallicity, especially as it is near the edge of the existing library.

\subsection{Stellar classification from LAMOST}
\label{sec:lamost}
The Large Sky Area Multi-Object Fibre Spectroscopic Telescope (LAMOST) is a 4-m telescope that uses 4000 fibers distributed over a 5\degr\ field of view to acquire low-resolution ($R\approx1800$) spectra in the optical (3700-9000\AA) band \citep{Cui2012}. TOI-5349 was observed on 27 December 2015 as part of a spectroscopic survey of the Galactic anticenter \citep{Deng2012,Yuan2015,Xiang2017} and the spectrum was made public as part of DR10v2.0\footnote{\url{https://www.lamost.org/dr10/v2.0/}} \citep{Wang2022}. 

LAMOST DR10 derives a spectral type, spectroscopic parameters and uncertainties (derived via cross-validation) using an empirical sample of 1365 high S/N M-dwarfs \citep{Du2021,Du2024}. The LAMOST stellar classification pipeline uses these empirical templates to identify molecular absorption features \citep[e.g., CaH, TiO; see][]{Lepine2007} and derives the spectral type of an M-dwarf with an accuracy of $\pm0.5$ subtypes \citep[][]{Zhong2015,Zhong2019}. From the LAMOST spectra, TOI-5349 is classified as an M$1\pm0.5$ dwarf with $T_{e}=3675\pm67$ K and $\log g_\star=4.71\pm0.07$. The spectroscopic parameters are consistent with the parameters derived from higher-resolution near-infrared spectra in Section \ref{sec:specmatch} that are presented in \autoref{tab:stellarparam}.

\subsection{Fitting the spectral energy distribution}
Similar to \cite{Canas2023}, we derive stellar parameters by modeling the spectral energy distribution (SED) using the \texttt{EXOFASTv2} analysis package \citep{Eastman2019}, which fits the observed magnitudes using predictions from the MIST model grids \citep{Dotter2016,Choi2016}. The SED fit used Gaussian priors on the (i) broadband photometry listed in \autoref{tab:stellarparam}, (ii) spectroscopic parameters from \texttt{HPF-SpecMatch}, and (iii) parallax measurements from Gaia DR3 \citep{GaiaCollaboration2022}. We placed an upper limit on the visual extinction ($A_V$) based on estimates of Galactic dust \citep{Green2019}. \texttt{EXOFASTv2} adopts the $R_{v}=3.1$ reddening law from \cite{Fitzpatrick1999} to calculate a visual magnitude extinction during the SED fit. \autoref{tab:stellarparam} contains the derived stellar parameters. TOI-5349 has a mass and radius of $M_\star=0.61 \pm 0.02~\mathrm{M_{\odot}}$ and $R_\star=0.58 \pm 0.02~\mathrm{R_{\odot}}$, respectively.

\subsection{Galactic kinematics}
We calculated the $UVW$ velocities in the barycentric frame using GALPY \citep{Bovy2015} to determine the probabilities of TOI-5349 belonging to the galactic thin disk, thick disk, or halo. Using the equations in \cite{Bensby2014}, we determine a 98.60\%\ probability that TOI-5349 belongs to the thin-disk population when compared to the halo or thick disk population (see \autoref{tab:stellarparam}).

\section{Data Analysis}\label{sec:datafit}
We performed a joint fit of the RV and photometry data using the \texttt{exoplanet} software package \citep{Foreman-Mackey2021}, which employs the No-U-Turn Sampler (NUTS) with the Hamiltonian Monte Carlo (HMC) algorithm in \texttt{PyMC3} \citep{pymc3} for posterior estimation due to its efficiency in exploring the parameter space of correlated, high dimensional posteriors by auto-tuning the step size and choosing the trajectory length via a U-turn criterion. The \texttt{exoplanet} package uses \texttt{starry} \citep{starry} package to model the photometry data, which applies a quadratic limb-darkening law for transit modeling. The limb-darkening priors were reparameterized following \citet{Kipping2013b} for a quadratic limb-darkening law, with each dataset assigned its own limb-darkening coefficients. The TESS photometry did not exhibit any out-of-transit variability, and we did not include a Gaussian process in our photometric model. The RV data were fitted using a Keplerian model with an eccentric orbit, incorporating an RV offset and a jitter term to account for any systematic differences between instruments and astrophysical factors that may be present. We first obtained the maximum a posteriori (MAP) solution, using random starting conditions from priors specified in \autoref{tab:5349par}. This MAP solution was then used as the initial conditions for the NUTS sampler. The final joint fit was sampled using 4 independent chains with 6,000 tuning steps and the convergence was assessed using the Gelman-Rubin statistic \citep[all parameters had $\hat{R}<1.01$;][]{Gelman1992}. The system parameters derived from the final joint fit are presented in \autoref{tab:5349par}. The best-fit transit model is shown in \autoref{fig:lc}, while the best-fit RV model is shown in \autoref{fig:rv}.

\begin{deluxetable*}{llcccccc}
\tabletypesize{\small}
\tablecaption{System parameters for TOI-5349 \label{tab:5349par}}
\tablehead{\colhead{~~~Parameter} &
\colhead{Units} &
\colhead{Prior$^a$} &
\multicolumn{5}{c}{Value} 
}
\startdata
\sidehead{HPF parameters:}
~~~Systemic velocity & $\gamma~\mathrm{(m~s^{-1})}$ & $\mathcal{N}(0,2\times10^4)$ & \multicolumn{5}{c}{$70 \pm 20$} \\
~~~RV Jitter & $\sigma_{RV}~\mathrm{(m~s^{-1})}$ & $\mathcal{U}(10^{-3},10^3)$ & \multicolumn{5}{c}{$50_{-10}^{+20}$}\\
\sidehead{MAROON-X (Blue)  parameters:}
~~~Systemic velocity & $\gamma~\mathrm{(m~s^{-1})}$ & $\mathcal{N}(0,2\times10^4)$ & \multicolumn{5}{c}{$19 \pm 3$} \\
~~~RV Jitter & $\sigma_{RV}~\mathrm{(m~s^{-1})}$ & $\mathcal{U}(10^{-3},10^3)$ & \multicolumn{5}{c}{$3_{-2}^{+4}$}\\
\sidehead{MAROON-X (Red) parameters:}
~~~Systemic velocity & $\gamma~\mathrm{(m~s^{-1})}$ & $\mathcal{N}(0,2\times10^4)$ & \multicolumn{5}{c}{$19 \pm 3$} \\
~~~RV Jitter & $\sigma_{RV}~\mathrm{(m~s^{-1})}$ & $\mathcal{U}(10^{-3},10^3)$ & \multicolumn{5}{c}{$3_{-2}^{+4}$}\\
\hline
\sidehead{System parameters:}
~~~Orbital period & $P$ (days)  & $\mathcal{N}(3.3,0.1)$ & \multicolumn{5}{c}{$3.317921\pm0.000002$}\\
~~~Time of mid-transit & $T_0$ (BJD\textsubscript{TDB}) & $\mathcal{N}(2459521.8,0.1)$ & \multicolumn{5}{c}{$2459521.8184 \pm 0.0005$}\\
~~~$\sqrt{e}\cos\omega_\star$ &  \nodata & $\mathcal{U}(-1,1)$ & \multicolumn{5}{c}{$0.10_{-0.08}^{+0.05}$}\\
~~~$\sqrt{e}\sin\omega_\star$ &  \nodata & $\mathcal{U}(-1,1)$ & \multicolumn{5}{c}{$-0.13_{-0.09}^{+0.14}$}\\
~~~Scaled radius & $R_{p}/R_{\star}$  & $\mathcal{U}(0,1)$ & \multicolumn{5}{c}{$0.161 \pm 0.002$}\\
~~~Impact parameter & $b$ & $\mathcal{U}(0,1)$ & \multicolumn{5}{c}{$0.51 \pm 0.04$}\\
~~~Mass & $M_{p}$ ($\mathrm{M_J}$)  & $\mathcal{U}(10^{-3},10^3)$  & \multicolumn{5}{c}{$0.40 \pm 0.02$}\\
~~~\nodata & $M_{p}$ ($\mathrm{M_\oplus}$)  & \nodata  & \multicolumn{5}{c}{$127.4_{-5.5}^{+5.8}$}\\
\hline
\sidehead{Derived parameters:}
~~~Semi-amplitude velocity & $K~\mathrm{(m~s^{-1})}$  & \nodata &   \multicolumn{5}{c}{$76.2\pm2.8$}\\
~~~Eccentricity & $e$  & \nodata & \multicolumn{5}{c}{$0.03^{+0.03}_{-0.02}$, $3\sigma<0.12$}\\
~~~Argument of periastron & $\omega_\star$ (degrees)  & \nodata& \multicolumn{5}{c}{$-50_{-20}^{+60}$}\\
~~~Scaled semi-major axis & $a/R_{\star}$  & \nodata & \multicolumn{5}{c}{$13.6 \pm 0.3$}\\
~~~Orbital inclination & $i$ (degrees) & \nodata & \multicolumn{5}{c}{$87.9 \pm 0.2$}\\
~~~Transit duration & $T_{14}$ (hours) & \nodata & \multicolumn{5}{c}{$1.95_{-0.05}^{+0.04}$}\\
~~~Radius & $R_{p}$  ($\mathrm{R_{J}}$)  & \nodata &  \multicolumn{5}{c}{$0.91 \pm 0.02$}\\
~~~\nodata & $R_{p}$  ($\mathrm{R_{\oplus}}$)  & \nodata &  \multicolumn{5}{c}{$10.2 \pm 0.3$}\\
~~~Surface gravity & $\log g_{p}$  (cgs)  & \nodata &  \multicolumn{5}{c}{$3.08 \pm 0.03$}\\
~~~Density & $\rho_{p}$  ($\mathrm{g~cm}^{-3}$)  & \nodata &  \multicolumn{5}{c}{$0.66 \pm 0.06$}\\
~~~Semi-major axis & $a$ (au)  & \nodata & \multicolumn{5}{c}{$0.0369 \pm 0.0004$}\\
~~~Average Incident flux & $\langle F \rangle$ ($S_\oplus$) & \nodata & \multicolumn{5}{c}{$44.5\pm3.7$}\\
~~~Equilibrium temperature\(^{b}\) & $T_{eq}$ (K) & \nodata & \multicolumn{5}{c}{$719\pm15$}\\
\enddata
\tablenotetext{a}{$\mathcal{N}(\mu,\sigma)$ is a Gaussian prior with a mean and standard deviation of $\mu$ and $\sigma$, respectively. $\mathcal{U}(A,B)$ is a uniform prior with lower and upper limits of $A$ and $B$, respectively.}
\tablenotetext{b}{The planet is assumed to be a blackbody.}
\end{deluxetable*}

\begin{figure*}
    \centering
    \includegraphics[width=\linewidth]{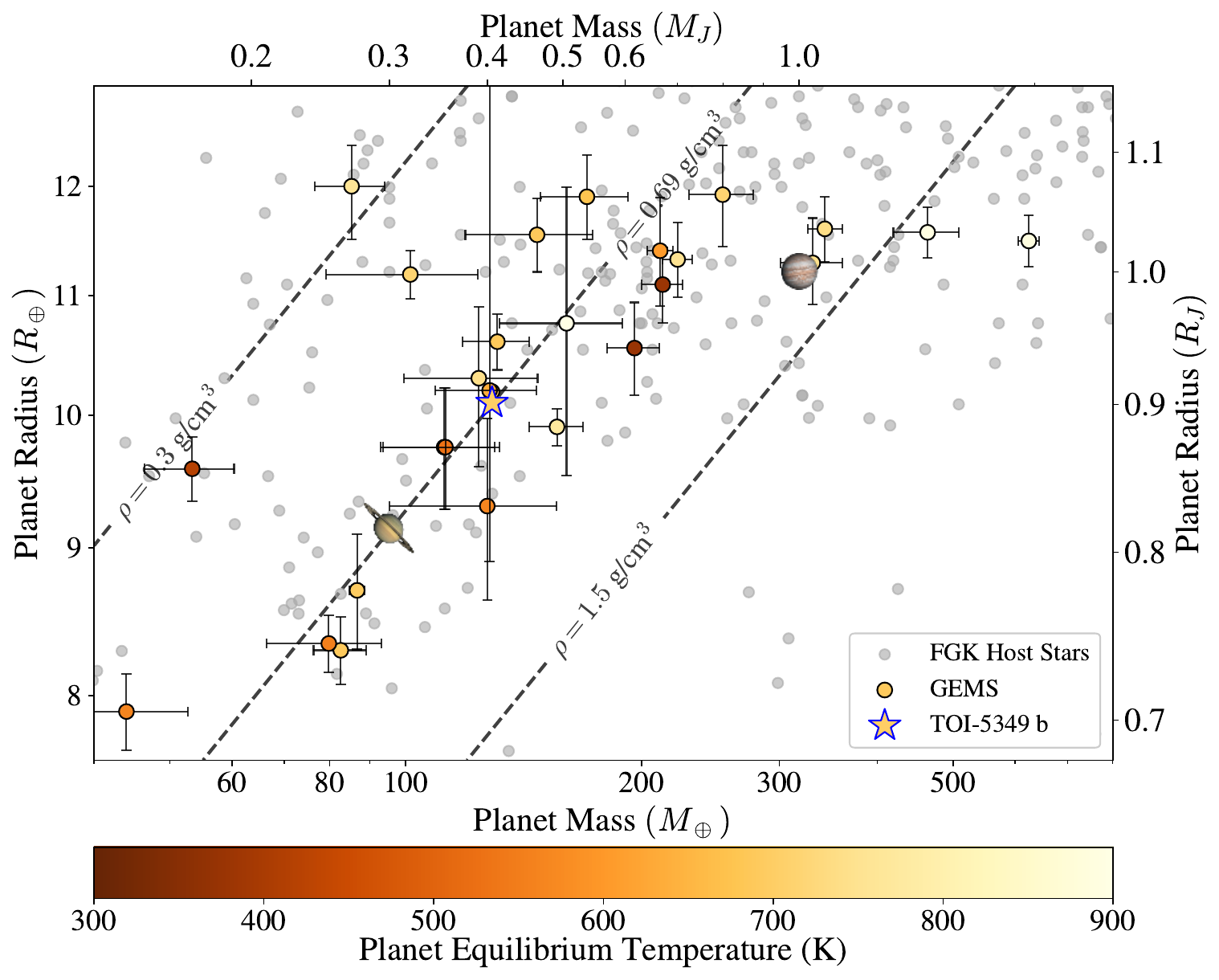}
    \caption{Planet mass as a function of planet radius, highlighting TOI-5349b (denoted by a star symbol) in the context of other GEMS planets. Planet masses are plotted in Earth masses ($M_\oplus$) and Earth radii ($R_\oplus$) as well as Jupiter masses $(M_J)$ and Jupiter radii $(R_J)$. TOI-5349b and GEMS targets are color-coded by their planet equilibrium temperature ($T_{\rm eq}$), with FGK-type host stars shown in gray circles for reference \citep{pscomp}. Density contours corresponding to $\rho = 0.3$, $0.69$, and $1.5$~g~cm$^{-3}$ are overlaid as dashed gray lines. Saturn and Jupiter are also shown for comparison. 
    \textit{Uncertainties in mass and radius for TOI-5349b are smaller than the marker size and thus not visible.}}
    \label{fig:mvr-plot}
\end{figure*}

\section{Discussion}
\label{sec:discussion}
\subsection{TOI-5349b in context of existing GEMS}
 We compare TOI-5349b with other well-characterized ($\geq 3\sigma$ mass and radius confidence) GEMS planets and the M-dwarf planet population. Using the NASA Exoplanet Archive \citep{Akeson2013}, we compiled a sample of GEMS with radii $\gtrsim 8~R_\oplus$, stellar effective temperatures $T_{\rm eff} < 4000$~K, mass and radius determinations with $\geq 3\sigma$ confidence from the planetary composite table \citep{pscomp}. Figure~\ref{fig:mvr-plot} shows the resulting mass-radius distribution, with density contours overlaid by dashed lines.
The GEMS sample ($N=30$) predominantly falls within a Saturn-like density regime ($0.69 \pm 0.0002~\mathrm{g~cm^{-3}}$), with densities between $0.4-0.9~\mathrm{g~cm^{-3}}$ and masses spanning $95-140~\mathrm{M_\oplus}$. TOI-5349b has a mass of $0.40 \pm 0.02~M_{\rm J}$ ($127.4_{-5.7}^{+5.9}~M_\oplus$) and a radius of $0.91 \pm 0.02~R_{\rm J}$ ($10.2 \pm 0.03~R_\oplus$), yielding a density of $\rho_p=0.66 \pm0.06~\mathrm{g~cm^{-3}}$ making it comparable to other Saturn-density GEMS such as TOI-5344b \citep{Hartman2023, Han2024}, TOI-5573b \citep{Fernandes2025}, TOI-5688 Ab \citep{Reji2025}, and TOI-6158b (O'Brien et al. in prep.) (See Table~\ref{tab:saturn_gems} for list of properties.)

TOI-5349b falls neatly along the pattern of Saturn-like density GEMS, demonstrating its consistency with other gas giants in this mass regime. Table~\ref{tab:saturn_gems} compares the properties of Saturn-like GEMS (Data from the NASA Exoplanet Archive was queried on 2025 June 20). Most exhibit masses of 110-130~$M_\oplus$, radii near 10~$R_\oplus$, and densities between 0.65 - 0.80~g~cm$^{-3}$. Scaled radii ($R_p/R_\star \sim 0.16$–0.22) and separations ($a/R_\star \sim 10$–16) are consistent, suggesting similar environments and potentially disk migration histories with TOI-5349b fitting squarely within these ranges. 

Recent work by \cite{Kanodia2025} found that when super-Jupiters ($\gtrsim 2~M_{\rm J}$) are excluded from a sample of transiting giant planets, the average masses of Jupiter-sized planets ($\geq 8~R_\oplus$) show no statistical difference between M-dwarfs and FGK hosts. The lack of mass differences suggests that these planets potentially have similar formation processes and initial conditions across different stellar masses, such as a possible minimum disk mass threshold. Metal-rich, high-mass disks appear to be the driving force for forming Saturn- and Jupiter-mass-like planets. 

The prevalence of Saturn-like masses and densities among the GEMS population may also be linked to the formation processes involved. The leading theory for the formation of gas giant planets is core accretion \citep{Pollack1996, Mizuno1980}, where dust grains in the protoplanetary disk begin clumping to form planetesimals until they reach a 10 ~$M_\oplus$ core. After the core reaches this critical mass, it can begin accreting gas and grow into a planet, however, this can also result in forming a range of planetary masses if there is insufficient material for runaway gas accretion. \cite{Helled2023A} proposed that runaway gas accretion is triggered around Saturn mass or $\sim$100~$M_\oplus$, marking the transition into gas giant planets. However, this transition is dependent on the exact formation history, stellar environment, and disk gas opacity. The onset of runaway gas accretion can be delayed or impeded through various processes, such as an intermediate phase of heavy-element accretion that can stall or suppress runaway accretion, causing the planet to remain at a Saturn-like mass if the gas disk disperses early \citep{Vazan2018, Muller2020}. This results in ``failed gas giants’’ which grew large cores but could not accrete the required gas to become Jupiter-mass planets. The similar composition and properties of TOI-5349b and other GEMS planets suggest that these planets may have formed through core-accretion under similar disk conditions where either runaway gas accretion may be limited or there may not be enough heavy element material to form Jupiter-mass objects. 

While core accretion remains the favored formation mechanism for gas giants, alternative processes such as gravitational instability \citep{Boss1997, Boss2006} may be responsible for the formation of some GEMS planets and should not be excluded. In the gravitational instability scenario, regions of a massive, cool protoplanetary disk become unstable and rapidly collapse to form giant planets, bypassing the slow core buildup required by core accretion. Previous studies suggest that gravitational instability may account for the most massive GEMS planets ($\gtrsim 4~M_{\rm J}$; \citealt{Helled2014, Hotnisky2024, Kanodia2025}), such as TOI-2379b \citep{Bryant2024} or TOI-6303b/TOI-6330b \citep{Hotnisky2024}, which show high planet-to-star mass ratios. However, for the Saturn-like GEMS, the more plausible formation pathway appears to be core accretion, possibly quenched before the onset of runaway gas accretion. Additional discoveries and well-characterized systems are still needed to further investigate the formation mechanisms across the GEMS population.

\subsection{Metallicity Correlations Among Gas Giants Orbiting M-dwarfs} 
\label{SubSec:met_sec}

The GEMS sample shows a strong tendency for orbiting metal-rich stars with the median metallicity of confirmed transiting GEMS hosts being [Fe/H] = +0.27 dex \citep{Han2024}, significantly higher than that of the typical field M-dwarf metallicity ([Fe/H] = -0.07 dex in the H-band and [Fe/H] = +0.07 dex in the K-band; \citealt{Gan2025}). This emerging trend mirrors the well-established planet–metallicity correlation observed for hot Jupiters around FGK stars, where giant planets are preferentially found around metal-rich hosts \citep{Gonzalez1997, Fischer&Valenti2005}.

\cite{Gan2025} observed a subset of 22 GEMS and derived metallicities using empirical relationships for spectroscopic observations with SpeX at near-infrared wavelengths \citep{Johnson2009, RojasAyala2010, Gaidos2014} and found that gas giants strongly favor metal-rich M-dwarfs compared to the broader field population. This finding reinforces the notion that metallicity is a significant factor in the formation and evolution of giants planets around low-mass stars. However, future studies using a consistent methodology to derive metallicities may provide more reliable and refined measurements, which may help further refine these correlations.
While current metallicity measurements are sufficiently accurate for detailed analyses, it is important to note that they are derived from a combination of spectroscopic and photometric methods, introducing inherent uncertainties. As such, the results should be interpreted with some caution. However, we note that TOI-5349 is confirmed metal-rich by both HPF-SpecMatch and LAMOST spectra, despite the limitations associated with their respective stellar libraries. We therefore adopt the quantitative metallicity value for TOI-5349 in our analysis while bearing this caveat in mind. 

\begin{figure*}
    \centering
    \includegraphics[width=\linewidth]{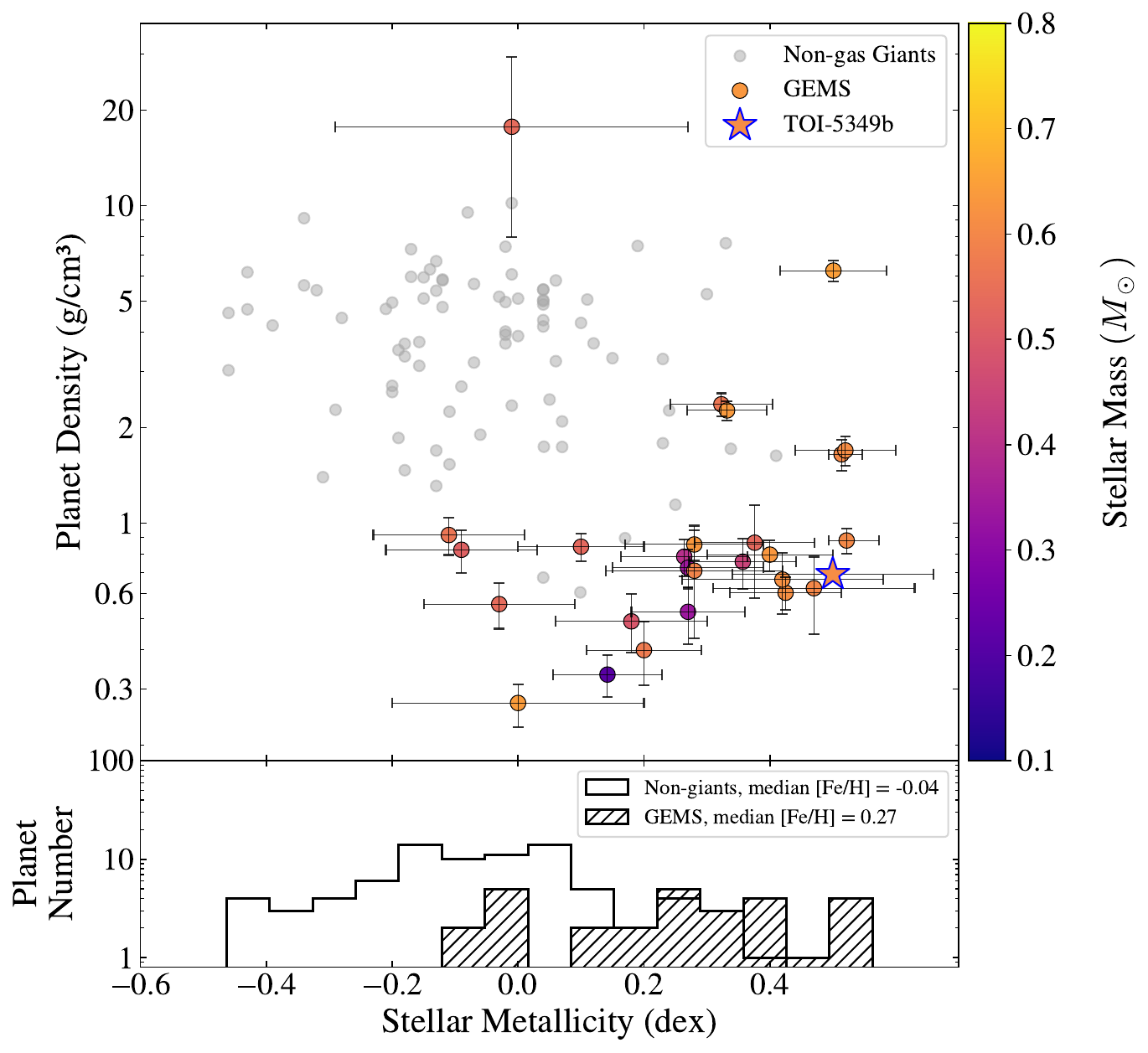}
    \caption{Planet density as a function of stellar metallicity for TOI-5349b (denoted by a star symbol), and other GEMS planets, which are color-coded by their host star mass ($M_\star$). Non-gas giant planets (e.g., super-Earths) are plotted in gray for comparison. Densities are plotted on a logarithmic scale in units of g~cm$^{-3}$, and stellar metallicities are given in [Fe/H] (dex).}
    \label{fig:mvd-plot}
\end{figure*}

TOI-5349b orbits a metal-rich M1-dwarf star with a metallicity of [Fe/H] = $+0.50 \pm 0.16$, confirmed both by HPF-SpecMatch and LAMOST (see \autoref{sec:specmatch} and \autoref{sec:lamost}). This places it among the most metal-rich GEMS hosts to date, second only to HATS-75b ([Fe/H] = $+0.52 \pm 0.03$) \citep{Jordan2022}. To place TOI-5349b in context within the broader metallicity trends of GEMS, we plot stellar metallicity as a function of planet density in \autoref{fig:mvd-plot}. We observe an emerging trend where low-density gas giants seem to favor higher stellar metallicities, with TOI-5349b aligning neatly with this pattern alongside other Saturn-like GEMS, further supporting this observed correlation. 

This pattern among GEMS systems suggests that core accretion is the dominant formation pathway, as high stellar metallicity is strongly correlated with gas giant occurrence. Metal-rich environments improve the solid surface density in protoplanetary disks, thereby accelerating the growth of planetary cores to the critical mass needed to initiate runaway gas accretion before the disk dissipates \citep{Kanodia2024}. Elevated metallicities increase disk opacity, which can slow radiative cooling during the gas accretion phase. This reduced cooling efficiency may delay or suppress runaway accretion, potentially stalling growth at Saturn-like masses rather than yielding Jupiter analogs \citep{Helled2014}. This mechanism may help explain why many GEMS planets exhibit masses and radii intermediate between those of Saturn and Jupiter.

\startlongtable
\begin{deluxetable*}{lccccccccc}
\tabletypesize{\footnotesize}
\tablenum{4}
\tablecaption{Properties of Saturn-like GEMS, sorted by planetary mass.\label{tab:saturn_gems}}
\tablehead{
\colhead{Planet Name} & \colhead{Orbital Period} & \colhead{Mass} &
\colhead{Radius} & \colhead{$T_{\rm eq}$} & \colhead{$S$} &
\colhead{$R_p/R_\star$} & \colhead{$a/R_\star$} & \colhead{Density} & \colhead{Reference} \\
 & (days) & ($\mathrm{M_\oplus}$) & ($R_\oplus$) & (K) &  ($S_\oplus$) & & & ($\mathrm{g~cm^{-3}}$)
}
\startdata
TOI-762 A b & 3.47 & 79.8 & 8.3 & 555 & 17.00 & 0.18 & 17.29 & 0.76 & 1 \\
TOI-3629 b & 3.94 & 82.6 & 8.3 & 690 & 39.00 & 0.13 & 15.40 & 0.80 & 2, 3 \\
TOI-4860 b & 1.52 & 86.7 & 8.7 & 694 & 42.50 & 0.22 & 11.00 & 0.75 & 4, 5 \\
TOI-5573 b & 8.80 & 112.0 & 9.8 & 528 & 12.92 & 0.150 & 25.74 & 0.66 & 6 \\
TOI-5688 A b & 2.95 & 124.0 & 10.3 & 742 & 50.3 & 0.164 & 12.5 & 0.61 & 7 \\
TOI-6158 b & 3.05 & 128.0 & 10.2 & 633 & 26.7 & 0.204 & 15.0 & 0.65 & 8 \\
\textbf{TOI-5349 b} & \textbf{3.32} & \textbf{128.9} & \textbf{10.2} & \textbf{700} & \textbf{44.5} & \textbf{0.161} & \textbf{13.6} & \textbf{0.66} & \textbf{This work} \\
TOI-5344 b & 3.79 & 131.0 & 10.6 & 689 & 38.0 & 0.1653 & 14.78 & 0.60 & 9, 10 \\
Kepler-45 b & 2.46 & 135.2 & 10.6 & 887 & 85.2 & 0.153 & 11.36 & 0.72 & 11 \\
HATS-75 b & 2.79 & 156.1 & 9.9 & 772 & 59.26 & 0.16 & 12.04 & 0.88 & 12 \\
\enddata
\tablerefs{
1: \citet{Hartman2024},
2: \citet{Canas2022},
3: \citet{Hartman2023},
4: \citet{Triaud2023},
5: \citet{Almenara2024},
6: \citet{Fernandes2025},
7: \citet{Reji2025},
8: O'Brien et al. in prep.,
9: \citet{Hartman2023},
10: \citet{Han2024},
11: \citet{Johnson2012}
12: \citet{Jordan2022},
}\end{deluxetable*}

\subsection{Feasibility of Atmospheric Studies in the GEMS Sample}
\begin{figure}
    \centering
    \includegraphics[width=\linewidth]{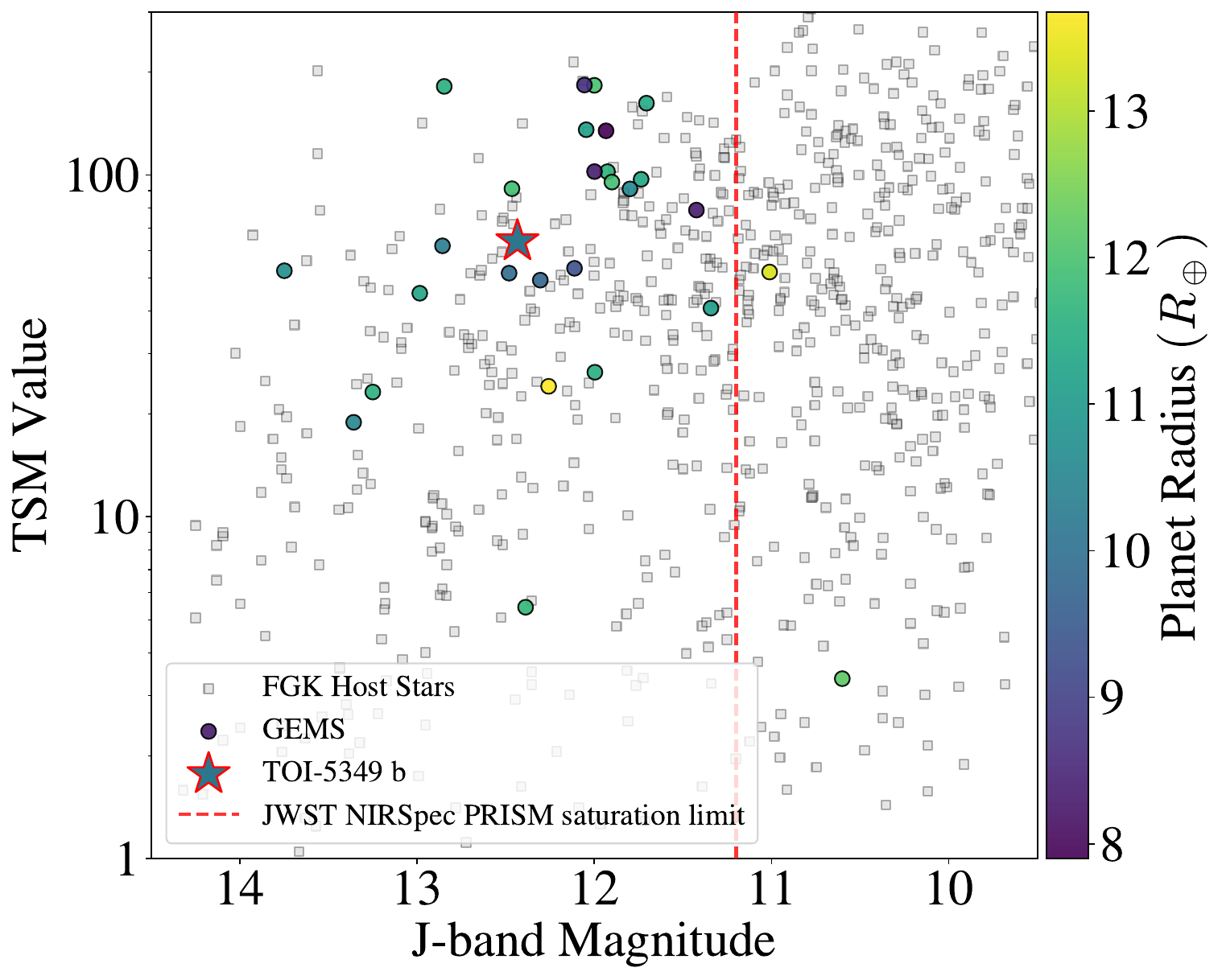}
    \caption{Transmission Spectroscopy Metric (TSM) as a function of host star J-band magnitude for the GEMS sample. The plot highlights how TSM varies with host star J-band magnitude, revealing a steep decline in transmission signal strength for dimmer stars. TOI-5349b lies at J = 12.43 with a TSM of $64.0 \pm 6.2$, which is well below the saturation threshold for transmission spectroscopy for an early M-dwarf \citep[$J\sim11.2$;][]{Birkmann2022}, making it a viable target for observations with JWST.}
    \label{fig:JbandTSM-plot}
\end{figure}

To further characterize the GEMS population and evaluate their potential for atmospheric observations, we examined the Transmission Spectroscopy Metric (TSM; \citealt{Kempton2018}) and host star brightness (J-band magnitude). We computed the TSM values for the Saturn-like GEMS sample, incorporating propagated uncertainties from planet radius, mass, stellar radius, equilibrium temperature, and J-band magnitude (see \autoref{fig:JbandTSM-plot}).

Across the population, the TSM values span a wide range, from as low as 3.4 to as high as $183.5 \pm 12.4$. TOI-5349b has a TSM of $64.0 \pm 6.2$, placing it near the median of the current sample. Additionally, TOI-5349b's moderate host star brightness (J = 12.43) avoids the saturation limits of JWST NIRSpec PRISM mode (J $\lesssim$ 11.2), making it well-suited for complete wavelength coverage observations spanning 0.6 - 5.3~$\mu$m without requiring alternative observing methods. Other favorable candidates include TOI-5344b ($91.1 \pm 12.2$), TOI-3629b ($79.0 \pm 10.5$), and HATS-75b ($51.6 \pm 5.4$), each lying above the typical detectability threshold for transmission spectroscopy. These targets present an opportunity to explore a more diverse set of Saturn-like GEMS orbiting more metal-rich early to mid M-dwarfs. 

Atmospheric observations of Saturn-like GEMS, such as TOI-5349b, can offer valuable constraints on formation and evolutionary pathways. These intermediate-mass planets occupy a regime in which atmospheric metallicity \citep{Guillot2023,Swain2024}, scale height, and cloud structure can diverge significantly from both hot Jupiters and Solar System gas giant planets. Recent transmission spectra of a similar GEMS planet, TOI-5205b, show an incredibly low atmospheric metallicity despite its host star being super-solar in metallicity \citep{Kanodia2023,Canas2025}. The transmission spectra of TOI-5205b reveal that the atmosphere is mainly dominated by stellar contamination from sunspots, along with detections of CH$_4$ and H$_2$ highlighting the complex relationship between stellar and planetary composition in these systems. One advantage for studying TOI-5349b is that it orbits a relatively inactive M1 dwarf, which minimizes stellar contamination that can complicate atmospheric retrievals. The TESS photometry and our transit modeling revealed no significant out-of-transit variability and no flares, suggesting limited short-period stellar activity that could complicate atmospheric retrievals. In addition, our spectroscopic diagnostics show no correlation between activity indicators (e.g., differential line width) and the measured RVs. While we cannot fully exclude low-level activity, the absence of strong activity signatures indicates that stellar contamination is unlikely to dominate transmission spectroscopy, making TOI-5349b a favorable target for follow-up atmospheric characterization and could help determine whether the unusual atmospheric properties seen in TOI-5205b represent a broader trend among GEMS planets or reflect unique formation and evolutionary processes.

Moreover, comparative studies of planets with similar bulk properties but differing host star metallicities and effective temperatures are particularly compelling. For instance, TOI-5349b and HATS-75b both orbit metal-rich M-dwarfs and fall within the same Saturn-like regime. HATS-75b is already being targeted under the JWST GO program 3171 \citep{jwstgo3171}, and future observations of TOI-5349b with JWST would be particularly valuable for expanding the sample of Saturn-like GEMS with measured transmission spectra.

\section{Conclusion}
\label{sec:conclusion}
We have presented the discovery and characterization of TOI-5349b, a Saturn-like gas giant ($M_p = 0.40\,M_J \approx 127.4_{-5.7}^{+5.9}, M_\oplus$, $\rho_p = 0.66\,\mathrm{g\,cm^{-3}}$) transiting a metal-rich M1 dwarf. With a radius of $0.90\,R_J$, TOI-5349b occupies the parameter space of the emerging population of Saturn-like GEMS with radii between $8-10~\mathrm{R_\oplus}$, masses between $40-140~\mathrm{M_\oplus}$, and densities typically between $0.4-0.9~\mathrm{g\,cm^{-3}}$.
The super-solar metallicity of the host star and Saturn-like density of the planet support formation through core accretion. TOI-5349b's host star is one of the most metal-rich M-dwarfs known to host a transiting giant, reinforcing the observed trend that GEMS planets form around metal-rich stars. These results add to the increasing evidence that higher dust-to-gas ratios in protoplanetary disks enable the timely formation of massive cores even around lower-mass stars.

In summary, TOI-5349b adds to the emerging picture of Saturn-like planets forming preferentially around metal-rich M-dwarfs, where these planets tend to form in high-metallicity environments and exhibit low densities, supporting core accretion as their dominant formation mechanism. TOI-5349b also reinforces the need to explore how metallicity, stellar mass, and disk conditions shape the occurrence and properties of gas giants in M-dwarf systems. As the GEMS survey continues to expand this sample, systems like TOI-5349b will serve as key benchmarks for refining planet formation and migration models in low-mass stellar environments.
\begin{acknowledgments}
We thank the anonymous referee for valuable feedback which has improved the quality of this manuscript. 
AS acknowledges support from the Simons Foundation Presidents Discretionary Fund.
CIC acknowledges support from NASA Headquarters through an appointment to the NASA Postdoctoral Program at the Goddard Space Flight Center, administered by ORAU through a contract with NASA. 
RF acknowledges support from the NASA Hubble Fellowship grant HST-HF2-51559.001-A awarded by the Space Telescope Science Institute, which is operated by the Association of Universities for Research in Astronomy, Inc., for NASA, under contract NAS5-26555.

Resources supporting this work were provided by the (i) NASA High-End Computing Program through the NASA Center for Climate Simulation (NCCS) at Goddard Space Flight Center 
(ii) Pennsylvania State University's Institute for Computational and Data Sciences' (ICDS) Roar supercomputer, and 
(iii) VANIR high performance computing cluster at the American Museum of Natural History (AMNH). This content is solely the responsibility of the authors and does not represent the views of NCCS, ICDS, or AMNH.

The Center for Exoplanets and Habitable Worlds is supported by the Pennsylvania State University, the Eberly College of Science, and the Pennsylvania Space Grant Consortium. 

We acknowledge support from NSF grants AST 1006676, AST 1126413, AST 1310875, AST 1310885, AST 2009554, AST 2009889, AST 2108512, AST 2108801 and the NASA Astrobiology Institute (NNA09DA76A) in our pursuit of precision RVs in the near-infrared. We acknowledge the support of the Heising-Simons Foundation through grant 2017-0494. These results are based on observations obtained with HPF on the HET. The HET is a joint project of the University of Texas at Austin, the Pennsylvania State University, Ludwig-Maximilians-Universit\"at M\"unchen, and Georg-August Universit\"at Gottingen. The HET is named in honor of its principal benefactors, William P. Hobby and Robert E. Eberly. The HET collaboration acknowledges the support and resources from the Texas Advanced Computing Center. We are grateful to the HET Resident Astronomers and Telescope Operators for their valuable assistance in gathering our HPF data.
We would like to acknowledge that the HET is built on Indigenous land. Moreover, we would like to acknowledge and pay our respects to the Carrizo \& Comecrudo, Coahuiltecan, Caddo, Tonkawa, Comanche, Lipan Apache, Alabama-Coushatta, Kickapoo, Tigua Pueblo, and all the American Indian and Indigenous Peoples and communities who have been or have become a part of these lands and territories in Texas, here on Turtle Island.

Some of the observations in this paper made use of the NN-EXPLORE Exoplanet and Stellar Speckle Imager (NESSI). NESSI was funded by the NASA Exoplanet Exploration Program and the NASA Ames Research Center. NESSI was built at the Ames Research Center by Steve B. Howell, Nic Scott, Elliott P. Horch, and Emmett Quigley.

Some of the observations were obtained using MAROON-X (GN-2023B-Q-104, PI: Kanodia) at the international Gemini Observatory, a program of NSF NOIRLab, which is managed by the Association of Universities for Research in Astronomy (AURA) under a cooperative agreement with the U.S. National Science Foundation on behalf of the Gemini Observatory partnership: the U.S. National Science Foundation (United States), National Research Council (Canada), Agencia Nacional de Investigaci\'{o}n y Desarrollo (Chile), Ministerio de Ciencia, Tecnolog\'{i}a e Innovaci\'{o}n (Argentina), Minist\'{e}rio da Ci\^{e}ncia, Tecnologia, Inova\c{c}\~{o}es e Comunica\c{c}\~{o}es (Brazil), and Korea Astronomy and Space Science Institute (Republic of Korea). 
This work was enabled by observations made from the Gemini North telescope, located within the Maunakea Science Reserve and adjacent to the summit of Maunakea. We are grateful for the privilege of observing the Universe from a place that is unique in both its astronomical quality and its cultural significance.

Some of the data presented in this paper were obtained from MAST at STScI. Support for MAST for non-HST data is provided by the NASA Office of Space Science via grant NNX09AF08G and by other grants and contracts.
This work includes data collected by the TESS mission, which are publicly available from MAST. Funding for the TESS mission is provided by the NASA Science Mission directorate. We acknowledge the use of public TOI Release data from pipelines at the TESS Science Office and at the TESS Science Processing Operations Center.
This research made use of the (i) NASA Exoplanet Archive, which is operated by Caltech, under contract with NASA under the Exoplanet Exploration Program, (ii) SIMBAD database, operated at CDS, Strasbourg, France, (iii) NASA's Astrophysics Data System Bibliographic Services, (iv) NASA/IPAC Infrared Science Archive, which is funded by NASA and operated by the California Institute of Technology, and (v) data from 2MASS, a joint project of the University of Massachusetts and IPAC at Caltech, funded by NASA and the NSF.
The research was carried out, in part, at the Jet Propulsion Laboratory, California Institute of Technology, under a contract with the National Aeronautics and Space Administration (80NM0018D0004)

This work has used data from the European Space Agency (ESA) mission Gaia (\url{https://www.cosmos.esa.int/gaia}), processed by the Gaia Data Processing and Analysis Consortium (DPAC, \url{https://www.cosmos.esa.int/web/gaia/dpac/consortium}). Funding for the DPAC has been provided by national institutions, in particular the institutions participating in the Gaia Multilateral Agreement.

Some of the observations in this paper made use of the Guoshoujing Telescope (LAMOST) , a National Major Scientific Project built by the Chinese Academy of Sciences. Funding for the project has been provided by the National Development and Reform Commission. LAMOST is operated and managed by the National Astronomical Observatories, Chinese Academy of Sciences.
\end{acknowledgments}

\facilities{Exoplanet Archive, Gaia, Gemini:Gillett (MAROON-X), HET (HPF), LAMOST, MAST, RBO, TESS, TMO, WIYN (NESSI)} 
\software{
\texttt{astroquery} \citep{Ginsburg2019},
\texttt{astropy} \citep{astropy:2013, AstropyCollaboration2018, astropy:2022},
\texttt{barycorrpy} \citep{Kanodia2018}, 
\texttt{EXOFASTv2} \citep{Eastman2019},
\texttt{exoplanet} \citep{Foreman-Mackey2021},
\texttt{HPF-SERVAL} \citep{Stefansson2023},
\texttt{HPF-SpecMatch} \citep{Stefansson2020},
\texttt{lightkurve} \citep{LightkurveCollaboration2018},
\texttt{matplotlib} \citep{Hunter2007},
\texttt{numpy} \citep{vanderWalt2011},
\texttt{pandas} \citep{McKinney2010},
\texttt{PyMC3} \citep{pymc3},
\texttt{scipy} \citep{Virtanen2020},
\texttt{starry} \citep{starry},
\texttt{telfit} \citep{Gullikson2014},
\texttt{tglc} \citep{Han2023}
}

\bibliography{refs}{}
\bibliographystyle{aasjournal}

\end{document}